\documentclass[prl,twocolumn,amsmath,amssymb]{revtex4-1}
\usepackage{graphicx}% Include figure files
\usepackage{dcolumn}% Align table columns on decimal point
\usepackage{bm}% bold math
\usepackage{rotating}% provides a rotate environment
\usepackage{bbding}

\def\hu{\raisebox{0.1ex}{\begin{turn}{0}\hbox{\hskip -0.5em \setbox0\hbox{\raisebox{-0.55ex}{\Ellipse}}
    \rlap{\hskip 0.57em $\circ$}\box0}\end{turn}}\hskip 0.4em}
\def\hd{\raisebox{0.1ex}{\begin{turn}{0}\hbox{\hskip -0.5em \setbox0\hbox{\raisebox{-0.55ex}{\Ellipse}}
    \rlap{\hskip -0.05em $\circ$}\box0}\end{turn}}\hskip 0.4em}
\def\vu{\raisebox{0.1ex}{\begin{turn}{90}\hbox{\hskip -0.5em \setbox0\hbox{\raisebox{-0.55ex}{\Ellipse}}
    \rlap{\hskip 0.57em $\circ$}\box0}\end{turn}}\hskip 0.4em}
\def\vd{\raisebox{0.1ex}{\begin{turn}{90}\hbox{\hskip -0.5em \setbox0\hbox{\raisebox{-0.55ex}{\Ellipse}}
    \rlap{\hskip -0.05em $\circ$}\box0}\end{turn}}\hskip 0.4em}

\begin{document}

\title{Atomic structure of Mn wires on Si(001) resolved by scanning tunneling microscopy}

\author {A. Fuhrer}
\email{afu@zurich.ibm.com}
\author {F. J. Rue\ss}
\author{N. Moll}
\author {A. Curioni}
\author {D. Widmer}

\affiliation{IBM Research - Zurich,
S\"aumerstrasse 4, 8803 R\"uschlikon, Switzerland}

\begin{abstract}
At submonolayer coverage, Mn forms atomic wires on the Si(001) surface oriented perpendicular to the underlying Si dimer rows. While many other elements form symmetric dimer wires at room temperature, we show that Mn wires have an asymmetric appearance and pin the Si dimers nearby. We find that an atomic configuration with a Mn trimer unit cell can explain these observations due to the interplay between the Si dimer buckling phase near the wire and the orientation of the Mn trimer. We study the resulting four wire configurations in detail using high-resolution scanning tunneling microscopy (STM) imaging and compare our findings with STM images simulated by density functional theory. 
\end{abstract}


\maketitle
\begin{figure}[btp]
\includegraphics[width=80mm]{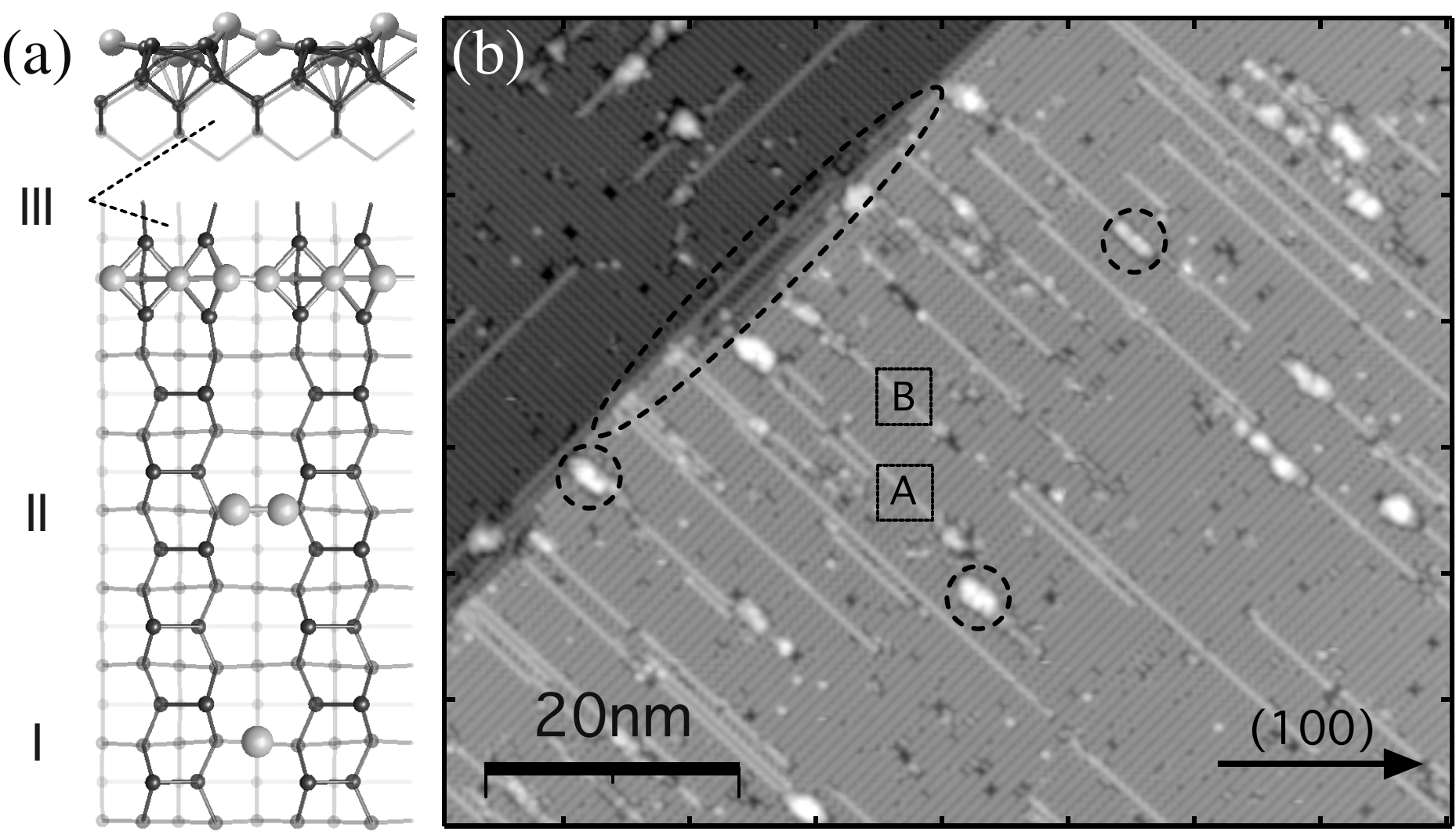}
\caption{\label{fig1}(a) Top and side view of the trimer wire structure (III) with an atom in the hollow site centered between two dimers of a dimer row. Mn atom positions for a dimer wire (II) and a wire with one Mn atom at the cave site (I) (we use the same terminology as in Ref. \cite{Sena11AA}). (b) STM image taken at $I_t=0.15$~nA and $V_s=-2.0$~V. Ellipse marks accumulation of Mn wires (bright lines) at Si step edge. Circles denote Mn clusters. }
\end{figure}

The large magnetic moment related to a half-filled $d$-shell renders Mn atoms attractive building blocks for fascinating magnetic nanostructures \cite{Hirjibehedin06AA,Sessoli93AA,Yakunin04AA}.
Tunable ferromagnetism in Mn-doped semiconductors has been achieved for GaAs, InAs and Ge \cite{Dietl02AA,Park02AB,Jamet06AA,Xiu10AB}. For silicon this effort has not been as successful because of strong segregation and the interstitial diffusion of Mn in the Si crystal during overgrowth or annealing, even though Mn implanted Si samples exhibit very high Curie temperatures \cite{Bolduc05AA}. 
During submonolayer deposition at room temperature (RT) however, Group III (Al, Ga, In),  Group IV (Sn, Pb) and a few other metals (e.g. Sb and Mg) are known to form 1D wires perpendicular to the Si dimer rows on the Si(100)-2$\times$1 surface \cite{Nogami91AA,Itoh93AA,Evans99AA,Dong98AB,Glueckstein98AA,Baski91AA,Itoh94AA,Nogami91AB,Kubo00AA}. These wires consist of metal atoms in the parallel-dimer configuration linking up to form atomic chains \cite{Northrup91AA,Brocks93AA,Bowler04AA}. 
Recent experiments show that similar wire formation occurs for Mn \cite{Liu08AA}. Density functional theory (DFT) calculations proposed several possible structures for these wires \cite{Sena11AA,Wang10AA,Niu11AA} with one, two or three Mn atoms per Si dimer row [see Fig.\;\ref{fig1}(a)]. %Albao05AA, Glueckstein98AA, Nogami91AA, Evans99AA

We show that high-resolution STM imaging at RT reveals a unique appearance of Mn wires. In contrast to other metal wires, their signature is characterized by the pinning of nearby Si dimers in addition to two independent asymmetric realizations of the Mn wire. We find that only an \emph{extended} trimer model can explain our findings and identify a total of four wire configurations based on the relative orientation of the Si dimers and an asymmetric Mn trimer unit cell. We confirm our model using a sequence of STM images, where we observe sequential changes of the Si dimer and Mn trimer configuration. 
Our results are important for the successful integration of Mn atoms into future silicon spintronics devices.

\begin{figure}[tbp]
\includegraphics[width=80mm]{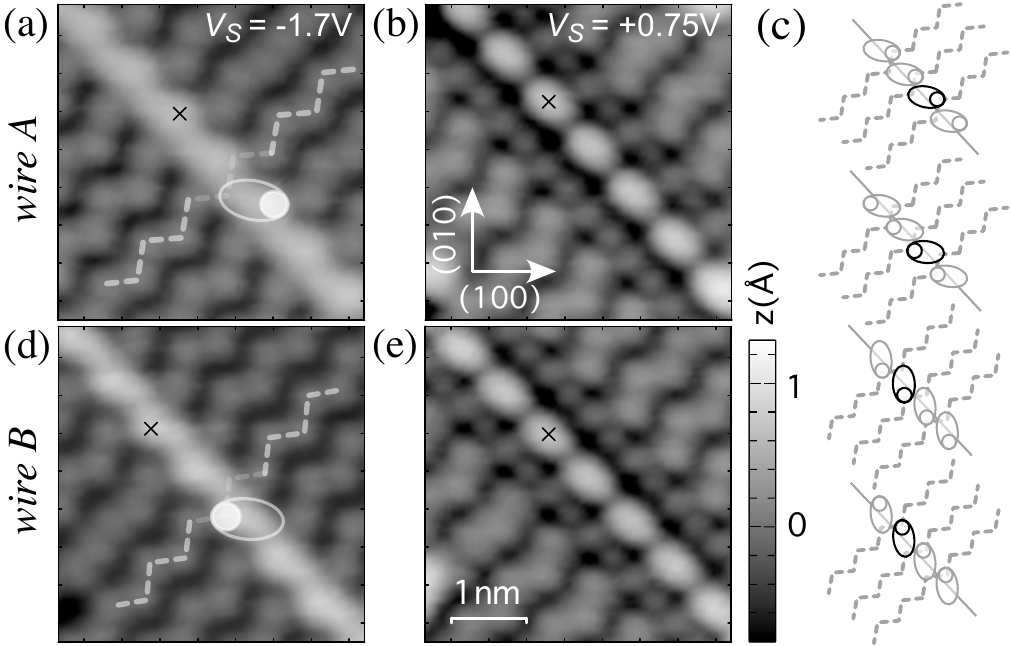}
\caption{\label{fig2} STM images of two wires A (a)+(b) and B (d)+(e). Tunneling parameters were $I_t=0.2$~nA and $V_s=-1.7$~V (a)+(d), $-0.75$~V (b)+(e). The four wire types that occur in our experiments are shown schematically in (c).}
\end{figure}

The filled-state STM image in Fig.\;\ref{fig1}(b) shows Mn wires near a step edge of the Si(001) surface. Here, about a tenth of a monolayer of Mn was deposited at RT with a rate of 55~pm/min. The wires, with typical lengths ranging from 5~nm to 50~nm, frequently nucleate at defect sites or step edges, where clustering of Mn wires is observed (dashed ellipse). This reflects the high mobility of Mn on the Si(001) surface at RT \cite{Liu08AA}. Even at these low-temperature growth conditions, a number of larger clusters are found (dashed circles). 
In the following, we take a closer look at two wires, denoted A and B in Fig.\;\ref{fig1}(b) and shown in higher resolution in Fig.\;\ref{fig2}(a) and (d). Interestingly, the two wires have a different appearance even though the reconstruction of the Si surface looks similar in the two images. Two things may be noted: First, the Si dimer buckling is pinned near the Mn wire. It has been argued \cite{Cho93AA} that strain along the dimer rows can immobilize the Si dimers, which otherwise move quickly back and forth giving the averaged appearance of the well-known Si 2$\times$1 reconstruction at RT. From the decay of the apparent dimer buckling (not shown), we estimate that the strain originates at the wire and decays over a distance of about $5\;$nm.
The second observation is that the Mn wires in filled state images have an asymmetric shape with respect to the wire axis with a sawtooth-like appearance. In addition to this asymmetry, Mn subunits between two Si dimer rows are also tilted. In Fig.\;\ref{fig2}(a) and (d) we highlight this tilt by an open ellipse in which the filled circle marks the side with the toothed edge. The phase of the Si dimer buckling stays continuous across the Mn wire, as indicated by the dashed lines in Fig.\;\ref{fig2}. The appearance of the wires does not depend on the wire length, but can sometimes change abruptly because of nearby defects. Experimentally, we find four types of wires as indicated in Fig.\;\ref{fig2}(c) with equal probability. We reference these in the text using the symbols \hu, \hd, \vd and \vu. The tilt of the wire subunits is always the same for a given Si dimer buckling phase (see dashed lines). For a given tilt, the sawtooth appearance is found to have two realizations with a rotation by $180^\circ$ connecting the two. 

The two panels (b) and (e) in Fig.\;\ref{fig2} show the asymmetry of the two wires when imaged at small empty state bias voltage $V_s=+0.75\;$V. The Si dimer atoms closest to the Mn wire are again hidden by the wire contrast. However, the two next-nearest Si dimers exhibit an atom-like appearance similar to that of normal Si dimers when the empty dangling bond states are imaged at more positive $V_s$. The asymmetry shows up as a dark gap on the side of the wire that was smooth in the filled-state images. The tilt of the wire subunits is not clearly visible and the highest apparent point (marked by $\times$) has moved away from the toothed edge towards the cave site.

\begin{figure*}[th!]
\includegraphics[width=165mm]{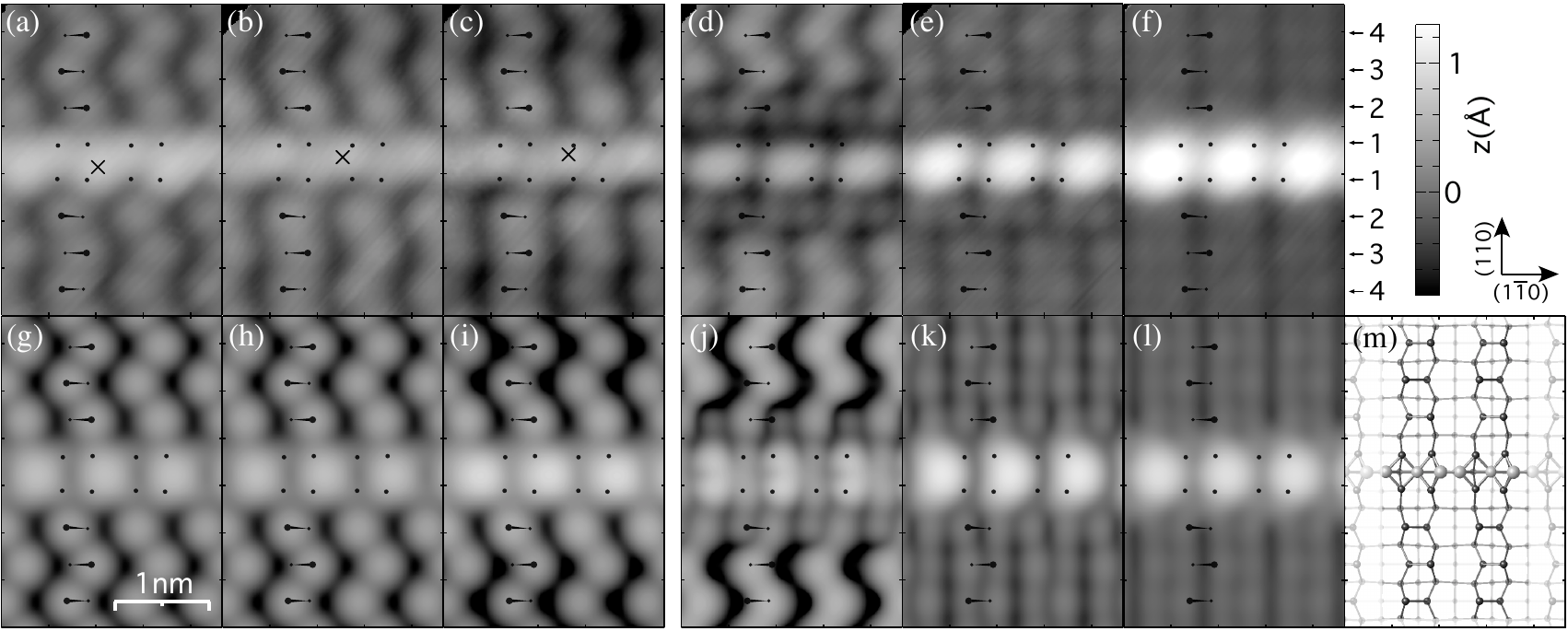}
\caption{\label{fig3} (a)-(f) Drift corrected STM images as a function of bias (a) $V_s=-1.75$\;V, (b) $-1.00$\;V, (c) $-0.65$\;V, (d) $+0.65$\;V, (e) $+1.00$\;V and (f) $+1.75$\;V. The tunneling current was $I_t=74$~pA for (c)+(d) and $154$~pA for all other images. The calculated positions of the Si dimer atoms are marked by the black overlay. (g)-(l) Simulated STM images from DFT (g) $E_B=-1.50$\;eV, (h) $-0.75$\;eV, (i) $-0.40$\;eV, (j) $+0.50$\;eV, (k) $+0.85$\;eV and (l) $+1.60$\;eV. (m) Relaxed atomic positions of the 8$\times$4 supercell. Arrows to the right of panel (f) indicate the positions of the 1st to 4th Si dimers away from the Mn wire.}
\end{figure*}

Comparing this behavior to that of other metal wires on Si\; \cite{Glueckstein98AA, Nogami91AA, Evans99AA} or to initial stages of Si and Ge growth on Si(001) \cite{Qin97AA}, it is clear that Mn is quite unique in its asymmetric appearance. All other metal wires are attributed to the parallel dimer wire model [see II in Fig.\;\ref{fig1}(a)] and are usually explained by a formation mechanism first described for Al \cite{Brocks93AA}. For these wires, filled-state STM images taken under similar conditions look completely symmetric with respect to the wire axis, except that in some cases a buckling of the metal dimers along the wire is observed \cite{Dong01AA}. As far as we know, none of these metal wires induces significant pinning of the Si dimer buckling close to the wire when imaged at RT. This suggests that the Mn wire structure creates more strain along the Si dimer rows. DFT calculations have shown that the hollow site is the most stable adsorption site for a single Mn atom on the Si(001) surface \cite{Hortamani06AA, Wang10AA}.  For one, two or three Mn adatoms, it is favorable to have an atom in the hollow site before the parallel dimer sites near the cave site are occupied. This was suggested to result in a trimer wire structure as shown in Fig.\;\ref{fig1}(a) III \cite{Wang10AA}. The additional atom in the hollow site could explain the experimentally observed pinning of the Si dimers. It is more difficult to explain the observed asymmetric appearance of the Mn wire because all three proposed Mn configurations in Fig.\;\ref{fig1}(a) are symmetric with respect to the wire axis. However, together with the pinned buckling of the Si dimers, the trimer wire structure is the only one that clearly breaks the twofold rotational symmetry present in symmetric metal wires. A rotation by $180^\circ$ keeps the dimer buckling on the surface the same, but the configuration of the two buckled Mn atoms near the cave site is mirrored. We therefore focus further discussion on comparing simulated STM images of the trimer wire structure at various tip-sample biases with our experimental results to identify the source of the unique asymmetric appearance of the Mn wires. 

To this end, we carried out state-of-the-art spin-polarized DFT calculations \cite{Hohenberg64AA,Kohn65AA} employing the highly optimized CPMD code \cite{_cpmd_2010}. The PBE (Perdew--Burke--Ernzerhof) exchange-correlation functional \cite{Perdew96AA} was applied and {\em ab-initio} norm-conserving pseudopotentials \cite{Hamann89AA} were used. The semilocal pseudopotentials were further transformed into fully separable Kleinman--Bylander pseudopotentials \cite{Kleinman82AA}, with the $d$-potential chosen as the local potential for the Mn atoms, $p$ for Si atoms, and $s$ for H atoms. %The wavefunctions at $0\;$K were expanded into plane waves \cite{Ihm79AA} with a kinetic energy of up to 100 Ry. An 8$\times$4 super cell was used with a thickness of 7 Si layers, in which the top 5 layers with Mn atoms were relaxed until the forces were smaller than 1 pN.  
The electron density was calculated from a 2$\times$1 $k$-point grid. The calculated wavefunctions from the $\Gamma$-point were analytically extended in $z$-direction away from the surface and integrated over the bias window. Finally, the $z$-value of constant density surfaces was plotted on the same grayscale as in the experiment. 
The results are shown in Fig.\;\ref{fig3}, where the top row shows experimental constant-current images. Here, the sample bias voltage $V_s$ is varied from filled- to empty-state imaging from left (a) to right (f). The bottom row shows the corresponding simulated images taking into account a slightly smaller effective band gap in the calculations. The images show very good agreement for the Si surface where the change in contrast for the Si dimers in all images is accurately reproduced up to the 2nd Si dimer away from the Mn wire. The observed change in appearance of the 2nd Si dimers in the empty-state images at low bias [Fig.\;\ref{fig3}(d),(e),(j),(k)] is related to a charge-transfer effect from the wire to the 2nd Si dimers. The atom positions of these Si dimers [see Fig.\;\ref{fig3}(m)] exhibit the usual buckled behavior with a buckling angle of $18^\circ$. This is not significantly different from the $18.7^\circ$ found for the 4th dimer pair, consistent with the bare Si value of $19\pm1^\circ$ in the literature \cite{Qin97AA,Ramstad95AA}. Nevertheless, the appearance of the 2nd dimers in Fig.\;\ref{fig3}(d) and (j) is roughly that of the 3rd and 4th dimers in Fig.\;\ref{fig3}(e) and (k) taken at a bias which is $0.35\;$eV higher. This indicates higher filling of the Si dimer states near the wire. A similar shift in contrast was observed for Ge \cite{Qin97AA} and Al \cite{Itoh93AA} wires, even though in these cases the hollow site is not occupied by a wire atom.

If we look at the wire appearance itself, the most obvious discrepancy between experiment and simulation is found in Fig.\;\ref{fig3}(d) and (j). The experimental image shows a horizontally elongated contrast centered on the hollow site, whereas the simulation displays a double-lobed vertical contrast centered over the highest Mn atom. In addition, the filled-state images in the simulation show nearly no bias dependance, whereas the experimental images exhibit a shift of the highest apparent point in the wire subunits indicated by $\times$ in Figs.\;\ref{fig3}(a), (b) and (c). We might expect such deviations due to several factors. For one, the simulation is done at $T=0\;$K, whereas all experimental images are taken at RT. Additionally, our assumption of a flat density of states of the STM tip reduces the variations between simulated filled-state images, especially at large bias. The periodic boundary conditions in the simulation further limit the appearance of wire asymmetries because the supercell has to be dipole free and charge transfer from one Mn wire edge to the other may be suppressed. Moreover, even if pure DFT simulations are accurate enough to describe the interaction and atomic structure of Mn on Si surfaces \cite{Sena11AA,Wang10AA,Niu11AA,Hortamani06AA}, the contraction of partially occupied $d$-shells is often underestimated, resulting in a modified orbital hybridization and therefore distorted simulated STM images. Nevertheless, we show that a closer look at the relaxed atom positions allows us to identify the origin of the observed asymmetry. 

\begin{figure}[t]
\includegraphics[width=85mm]{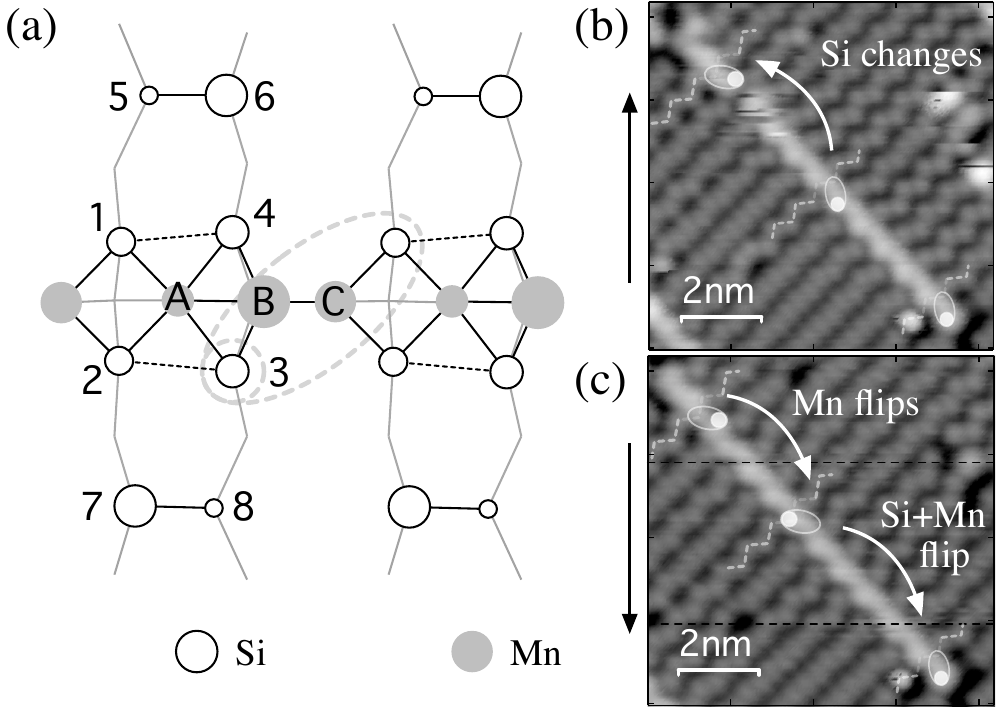}
\caption{\label{fig4} (a) Schematic of relaxed atom positions. Circle size indicates height of the atoms. The distortion of the Si atoms near the hollow site is exaggerated to illustrate the origin of the asymmetry. (b) and (c) Successive STM images of wire B before full relaxation of the Si dimer buckling. STM parameters were $I_t=0.2\;$nA, $V_s=-2.0\;$V (b) and $V_s=-1.7\;$V (c). Black arrows indicate the slow scan direction.}
\end{figure}

Despite significant differences between the empty-state images at low bias in Fig.\;\ref{fig3}(d) and (j), the simulation also reveals a slightly darker appearance of the 2nd dimers above the wire. This arises from a combination of two asymmetries, one due to the buckling of the Mn atoms along the wire, the other linked to the pinned Si dimers near the wire. The situation is shown schematically in Fig.\;\ref{fig4}(a). The two Si atoms (3,4) close to the highest Mn atom (B) are found to be 7\% further apart than Si atoms 1 and 2. Bond lengths 1--4 and 2--3 (see dashed lines) are the same on both sides and, with $3.1\;$\AA\ , about 30\% longer than that of normal Si dimers. The bond buckling angle is $3.4^\circ$ for bond 1--4 and $1.8^\circ$ for bond 2--3 such that Si atoms 3 and 4 are slightly higher than 1 and 2. Distance 2-7 is $4\;$\AA\ and identical to 3--8. In contrast to this, 4--6 is 3\% smaller and 5--1 is 3\% larger. This difference in proximity explains the higher contrast between the two atoms of the 2nd dimer (5,6) above the wire and the lower contrast for the 2nd dimer atoms below the wire (7,8) in the simulation. Furthermore, it leads to an overall darker appearance of the 2nd dimers above the wire, in agreement with the experiment. 

The mechanism for the asymmetric appearance relies on the relative orientation of the pinned Si dimer buckling and the buckling direction of the two Mn atoms near the cave site. Following our previous notation, the wire in Fig.\;\ref{fig4}(a) is a \hd wire and, together with \vu, has the highest Mn atom to the left of the cave site. Similarly, \hu and \vd have the highest Mn atom to the right of the cave site.
Fig.\;\ref{fig4}(b) and (c) show an example to highlight how the buckling directions of Si and Mn determine the appearance and atomic structure of the Mn wire as evidenced in sequential STM scans of wire B done shortly after Mn deposition. A rearrangement of Mn atoms during these scans changed the local environment sufficiently so that both the Si dimer buckling phase and the buckling direction of the Mn atoms along the wire was modified. Starting at the bottom in Fig.\;\ref{fig4}(b), the wire initially has a \vd appearance. In the top part, imaging shows instability followed by a switch to \hu for the last two wire subunits. This arises from a change in the buckling phase of the surrounding Si dimers. Scanning back down in Fig\;\ref{fig4}(c), the Si dimer buckling phase is now continuous over almost the entire image, but the apparent wire asymmetry still changes. This points to a switch of the Mn dimer buckling between the top \hu and the middle \hd part of the wire. Both the Si buckling phase and the Mn atom buckling change again at the bottom end of the wire, where two subunits in the \vd configuration remain. A further rearrangement of Mn atoms at the lower end of the wire finally converted almost the entire wire to \hd, as presented in Fig.\;\ref{fig2}(d). The two topmost subunits of the wire remained pinned in \hu and made scanning in that area unstable (not shown). This can be explained by a pinning of the Mn trimer orientation in opposite directions at the two ends of the wire, leaving an unstable boundary that is not present in other wires even though the Si dimer buckling phase is the same along the entire wire.

In conclusion, the unique asymmetric appearance of Mn atom wires on the Si(001) agrees best with an \emph{extended} trimer model. The asymmetry arises from the interplay between the orientation of the Si dimer buckling and the buckling of the Mn atoms near the cave site. Although many features seen experimentally could be reproduced in the simulated STM images, more refined models are needed to reproduce the exact appearance of the observed asymmetry. This is of particular interest in light of recent predictions of exciting spin and transport properties of Mn wires  \cite{Niu11AA}, which, if proven to be correct, may open the door to develop novel atomic-scale spintronic devices in silicon.

The authors acknowledge fruitful discussions with Rolf Allenspach and Gian Salis.
This research was supported by the Swiss National Science Foundation through the National Centre of Competence in Research Quantum Science and Technology.


\begin{thebibliography}{39}
\expandafter\ifx\csname natexlab\endcsname\relax\def\natexlab#1{#1}\fi
\expandafter\ifx\csname bibnamefont\endcsname\relax
  \def\bibnamefont#1{#1}\fi
\expandafter\ifx\csname bibfnamefont\endcsname\relax
  \def\bibfnamefont#1{#1}\fi
\expandafter\ifx\csname citenamefont\endcsname\relax
  \def\citenamefont#1{#1}\fi
\expandafter\ifx\csname url\endcsname\relax
  \def\url#1{\texttt{#1}}\fi
\expandafter\ifx\csname urlprefix\endcsname\relax\def\urlprefix{URL }\fi
\providecommand{\bibinfo}[2]{#2}
\providecommand{\eprint}[2][]{\url{#2}}

\bibitem[{\citenamefont{Hirjibehedin et~al.}(2006)\citenamefont{Hirjibehedin,
  Lutz, and Heinrich}}]{Hirjibehedin06AA}
\bibinfo{author}{\bibfnamefont{C.~F.} \bibnamefont{Hirjibehedin}},
  \bibinfo{author}{\bibfnamefont{C.~P.} \bibnamefont{Lutz}}, \bibnamefont{and}
  \bibinfo{author}{\bibfnamefont{A.~J.} \bibnamefont{Heinrich}},
  \bibinfo{journal}{Science} \textbf{\bibinfo{volume}{312}},
  \bibinfo{pages}{1021} (\bibinfo{year}{2006}).

\bibitem[{\citenamefont{Sessoli et~al.}(1993)\citenamefont{Sessoli, Gatteschi,
  Caneschi, and Novak}}]{Sessoli93AA}
\bibinfo{author}{\bibfnamefont{R.}~\bibnamefont{Sessoli}},
  \bibinfo{author}{\bibfnamefont{D.}~\bibnamefont{Gatteschi}},
  \bibinfo{author}{\bibfnamefont{A.}~\bibnamefont{Caneschi}}, \bibnamefont{and}
  \bibinfo{author}{\bibfnamefont{M.~A.} \bibnamefont{Novak}},
  \bibinfo{journal}{Nature} \textbf{\bibinfo{volume}{365}},
  \bibinfo{pages}{141} (\bibinfo{year}{1993}).

\bibitem[{\citenamefont{Yakunin et~al.}(2004)\citenamefont{Yakunin, Silov,
  Koenraad, Wolter, Van~Roy, De~Boeck, Tang, and Flatt\'e}}]{Yakunin04AA}
\bibinfo{author}{\bibfnamefont{A.~M.} \bibnamefont{Yakunin}},
  \bibinfo{author}{\bibfnamefont{A.~Y.} \bibnamefont{Silov}},
  \bibinfo{author}{\bibfnamefont{P.~M.} \bibnamefont{Koenraad}},
  \bibinfo{author}{\bibfnamefont{J.~H.} \bibnamefont{Wolter}},
  \bibinfo{author}{\bibfnamefont{W.}~\bibnamefont{Van~Roy}},
  \bibinfo{author}{\bibfnamefont{J.}~\bibnamefont{De~Boeck}},
  \bibinfo{author}{\bibfnamefont{J.-M.} \bibnamefont{Tang}}, \bibnamefont{and}
  \bibinfo{author}{\bibfnamefont{M.~E.} \bibnamefont{Flatt\'e}},
  \bibinfo{journal}{Phys. Rev. Lett.} \textbf{\bibinfo{volume}{92}},
  \bibinfo{pages}{216806} (\bibinfo{year}{2004}).

\bibitem[{\citenamefont{Dietl}(2002)}]{Dietl02AA}
\bibinfo{author}{\bibfnamefont{T.}~\bibnamefont{Dietl}},
  \bibinfo{journal}{Semicond. Sci. Technol.} \textbf{\bibinfo{volume}{17}},
  \bibinfo{pages}{377} (\bibinfo{year}{2002}).

\bibitem[{\citenamefont{Park et~al.}(2002)\citenamefont{Park, Hanbicki, Erwin,
  Helberg, Sullivan, Mattson, Ambrose, Wilson, Spanos, and Jonker}}]{Park02AB}
\bibinfo{author}{\bibfnamefont{Y.}~\bibnamefont{Park}},
  \bibinfo{author}{\bibfnamefont{A.}~\bibnamefont{Hanbicki}},
  \bibinfo{author}{\bibfnamefont{S.}~\bibnamefont{Erwin}},
  \bibinfo{author}{\bibfnamefont{C.}~\bibnamefont{Helberg}},
  \bibinfo{author}{\bibfnamefont{J.}~\bibnamefont{Sullivan}},
  \bibinfo{author}{\bibfnamefont{J.}~\bibnamefont{Mattson}},
  \bibinfo{author}{\bibfnamefont{T.~F.} \bibnamefont{Ambrose}},
  \bibinfo{author}{\bibfnamefont{A.}~\bibnamefont{Wilson}},
  \bibinfo{author}{\bibfnamefont{G.}~\bibnamefont{Spanos}}, \bibnamefont{and}
  \bibinfo{author}{\bibfnamefont{B.~T.} \bibnamefont{Jonker}},
  \bibinfo{journal}{Science} \textbf{\bibinfo{volume}{295}},
  \bibinfo{pages}{651} (\bibinfo{year}{2002}).

\bibitem[{\citenamefont{Jamet et~al.}(2006)\citenamefont{Jamet, Barski,
  Devilliers, Poydenot, Dujardin, Bayle-Guillemaud, Rothman, Bellet-Amalric,
  Marty, Cibert et~al.}}]{Jamet06AA}
\bibinfo{author}{\bibfnamefont{M.}~\bibnamefont{Jamet}},
  \bibinfo{author}{\bibfnamefont{A.}~\bibnamefont{Barski}},
  \bibinfo{author}{\bibfnamefont{T.}~\bibnamefont{Devilliers}},
  \bibinfo{author}{\bibfnamefont{V.}~\bibnamefont{Poydenot}},
  \bibinfo{author}{\bibfnamefont{R.}~\bibnamefont{Dujardin}},
  \bibinfo{author}{\bibfnamefont{P.}~\bibnamefont{Bayle-Guillemaud}},
  \bibinfo{author}{\bibfnamefont{J.}~\bibnamefont{Rothman}},
  \bibinfo{author}{\bibfnamefont{E.}~\bibnamefont{Bellet-Amalric}},
  \bibinfo{author}{\bibfnamefont{A.}~\bibnamefont{Marty}},
  \bibinfo{author}{\bibfnamefont{J.}~\bibnamefont{Cibert}},
  \bibnamefont{et~al.}, \bibinfo{journal}{Science}
  \textbf{\bibinfo{volume}{5}}, \bibinfo{pages}{653} (\bibinfo{year}{2006}).

\bibitem[{\citenamefont{Xiu et~al.}(2010)\citenamefont{Xiu, Wang, Kim, Hong,
  Tang, Jacob, Zou, and Wang}}]{Xiu10AB}
\bibinfo{author}{\bibfnamefont{F.}~\bibnamefont{Xiu}},
  \bibinfo{author}{\bibfnamefont{Y.}~\bibnamefont{Wang}},
  \bibinfo{author}{\bibfnamefont{J.}~\bibnamefont{Kim}},
  \bibinfo{author}{\bibfnamefont{A.}~\bibnamefont{Hong}},
  \bibinfo{author}{\bibfnamefont{J.}~\bibnamefont{Tang}},
  \bibinfo{author}{\bibfnamefont{A.~P.} \bibnamefont{Jacob}},
  \bibinfo{author}{\bibfnamefont{J.}~\bibnamefont{Zou}}, \bibnamefont{and}
  \bibinfo{author}{\bibfnamefont{K.~L.} \bibnamefont{Wang}},
  \bibinfo{journal}{Nature Materials} \textbf{\bibinfo{volume}{9}},
  \bibinfo{pages}{337} (\bibinfo{year}{2010}).

\bibitem[{\citenamefont{Bolduc et~al.}(2005)\citenamefont{Bolduc, Awo-Affouda,
  Stollenwerk, Huang, Ramos, Agnello, and LaBella}}]{Bolduc05AA}
\bibinfo{author}{\bibfnamefont{M.}~\bibnamefont{Bolduc}},
  \bibinfo{author}{\bibfnamefont{C.}~\bibnamefont{Awo-Affouda}},
  \bibinfo{author}{\bibfnamefont{A.}~\bibnamefont{Stollenwerk}},
  \bibinfo{author}{\bibfnamefont{M.~B.} \bibnamefont{Huang}},
  \bibinfo{author}{\bibfnamefont{F.~G.} \bibnamefont{Ramos}},
  \bibinfo{author}{\bibfnamefont{G.}~\bibnamefont{Agnello}}, \bibnamefont{and}
  \bibinfo{author}{\bibfnamefont{V.~P.} \bibnamefont{LaBella}},
  \bibinfo{journal}{Phys. Rev. B} \textbf{\bibinfo{volume}{71}},
  \bibinfo{pages}{033302} (\bibinfo{year}{2005}).

\bibitem[{\citenamefont{Nogami et~al.}(1991{\natexlab{a}})\citenamefont{Nogami,
  Baski, and Quate}}]{Nogami91AA}
\bibinfo{author}{\bibfnamefont{J.}~\bibnamefont{Nogami}},
  \bibinfo{author}{\bibfnamefont{A.~A.} \bibnamefont{Baski}}, \bibnamefont{and}
  \bibinfo{author}{\bibfnamefont{C.~F.} \bibnamefont{Quate}},
  \bibinfo{journal}{Phys. Rev. B} \textbf{\bibinfo{volume}{44}},
  \bibinfo{pages}{1415} (\bibinfo{year}{1991}{\natexlab{a}}).

\bibitem[{\citenamefont{Itoh et~al.}(1993)\citenamefont{Itoh, Itoh, Schmid, and
  Ichinokawa}}]{Itoh93AA}
\bibinfo{author}{\bibfnamefont{H.}~\bibnamefont{Itoh}},
  \bibinfo{author}{\bibfnamefont{J.}~\bibnamefont{Itoh}},
  \bibinfo{author}{\bibfnamefont{A.}~\bibnamefont{Schmid}}, \bibnamefont{and}
  \bibinfo{author}{\bibfnamefont{T.}~\bibnamefont{Ichinokawa}},
  \bibinfo{journal}{Phys. Rev. B} \textbf{\bibinfo{volume}{48}},
  \bibinfo{pages}{14663} (\bibinfo{year}{1993}).

\bibitem[{\citenamefont{Evans and Nogami}(1999)}]{Evans99AA}
\bibinfo{author}{\bibfnamefont{M.~M.~R.} \bibnamefont{Evans}} \bibnamefont{and}
  \bibinfo{author}{\bibfnamefont{J.}~\bibnamefont{Nogami}},
  \bibinfo{journal}{Phys. Rev. B} \textbf{\bibinfo{volume}{59}},
  \bibinfo{pages}{7644} (\bibinfo{year}{1999}).

\bibitem[{\citenamefont{Dong et~al.}(1998)\citenamefont{Dong, Yakabe, Fujita,
  and Nejoh}}]{Dong98AB}
\bibinfo{author}{\bibfnamefont{Z.}~\bibnamefont{Dong}},
  \bibinfo{author}{\bibfnamefont{T.}~\bibnamefont{Yakabe}},
  \bibinfo{author}{\bibfnamefont{D.}~\bibnamefont{Fujita}}, \bibnamefont{and}
  \bibinfo{author}{\bibfnamefont{H.}~\bibnamefont{Nejoh}},
  \bibinfo{journal}{Ultramicroscopy} \textbf{\bibinfo{volume}{73}},
  \bibinfo{pages}{169} (\bibinfo{year}{1998}).

\bibitem[{\citenamefont{Glueckstein et~al.}(1998)\citenamefont{Glueckstein,
  Evans, and Nogami}}]{Glueckstein98AA}
\bibinfo{author}{\bibfnamefont{J.}~\bibnamefont{Glueckstein}},
  \bibinfo{author}{\bibfnamefont{M.}~\bibnamefont{Evans}}, \bibnamefont{and}
  \bibinfo{author}{\bibfnamefont{J.}~\bibnamefont{Nogami}},
  \bibinfo{journal}{Surf. Sci.} \textbf{\bibinfo{volume}{415}},
  \bibinfo{pages}{80} (\bibinfo{year}{1998}).

\bibitem[{\citenamefont{Baski et~al.}(1991{\natexlab{b}})\citenamefont{Baski,
  Quate, and Nogami}}]{Baski91AA}
\bibinfo{author}{\bibfnamefont{A.~A.} \bibnamefont{Baski}},
  \bibinfo{author}{\bibfnamefont{C.~F.} \bibnamefont{Quate}}, \bibnamefont{and}
  \bibinfo{author}{\bibfnamefont{J.}~\bibnamefont{Nogami}},
  \bibinfo{journal}{Phys. Rev. B} \textbf{\bibinfo{volume}{44}},
  \bibinfo{pages}{11167} (\bibinfo{year}{1991}{\natexlab{b}}).

\bibitem[{\citenamefont{Itoh et~al.}(1994)\citenamefont{Itoh, Tanabe, Winau,
  Schmid, and Ichinokawa}}]{Itoh94AA}
\bibinfo{author}{\bibfnamefont{H.}~\bibnamefont{Itoh}},
  \bibinfo{author}{\bibfnamefont{H.}~\bibnamefont{Tanabe}},
  \bibinfo{author}{\bibfnamefont{D.}~\bibnamefont{Winau}},
  \bibinfo{author}{\bibfnamefont{A.}~\bibnamefont{Schmid}}, \bibnamefont{and}
  \bibinfo{author}{\bibfnamefont{T.}~\bibnamefont{Ichinokawa}},
  \bibinfo{journal}{J. Vac. Sci. Technol. B} \textbf{\bibinfo{volume}{12}},
  \bibinfo{pages}{2086} (\bibinfo{year}{1994}).

\bibitem[{\citenamefont{Nogami et~al.}(1991{\natexlab{b}})\citenamefont{Nogami,
  Baski, and Quate}}]{Nogami91AB}
\bibinfo{author}{\bibfnamefont{J.}~\bibnamefont{Nogami}},
  \bibinfo{author}{\bibfnamefont{A.~A.} \bibnamefont{Baski}}, \bibnamefont{and}
  \bibinfo{author}{\bibfnamefont{C.~F.} \bibnamefont{Quate}},
  \bibinfo{journal}{Appl. Phys. Lett.} \textbf{\bibinfo{volume}{58}},
  \bibinfo{pages}{475} (\bibinfo{year}{1991}{\natexlab{b}}).

\bibitem[{\citenamefont{Kubo et~al.}(2000)\citenamefont{Kubo, Saranin, Zotov,
  Harada, Kobayashi, Yamaoka, Ryu, Katayama, and Oura}}]{Kubo00AA}
\bibinfo{author}{\bibfnamefont{O.}~\bibnamefont{Kubo}},
  \bibinfo{author}{\bibfnamefont{A.~A.} \bibnamefont{Saranin}},
  \bibinfo{author}{\bibfnamefont{A.~V.} \bibnamefont{Zotov}},
  \bibinfo{author}{\bibfnamefont{T.}~\bibnamefont{Harada}},
  \bibinfo{author}{\bibfnamefont{T.}~\bibnamefont{Kobayashi}},
  \bibinfo{author}{\bibfnamefont{N.}~\bibnamefont{Yamaoka}},
  \bibinfo{author}{\bibfnamefont{J.~T.} \bibnamefont{Ryu}},
  \bibinfo{author}{\bibfnamefont{M.}~\bibnamefont{Katayama}}, \bibnamefont{and}
  \bibinfo{author}{\bibfnamefont{K.}~\bibnamefont{Oura}},
  \bibinfo{journal}{Jpn. J. of App. Phys.} \textbf{\bibinfo{volume}{39}},
  \bibinfo{pages}{3740} (\bibinfo{year}{2000}).

\bibitem[{\citenamefont{Northrup et~al.}(1991)\citenamefont{Northrup, Schabel,
  Karlsson, and Uhrberg}}]{Northrup91AA}
\bibinfo{author}{\bibfnamefont{J.~E.} \bibnamefont{Northrup}},
  \bibinfo{author}{\bibfnamefont{M.~C.} \bibnamefont{Schabel}},
  \bibinfo{author}{\bibfnamefont{C.~J.} \bibnamefont{Karlsson}},
  \bibnamefont{and} \bibinfo{author}{\bibfnamefont{R.~I.~G.}
  \bibnamefont{Uhrberg}}, \bibinfo{journal}{Phys. Rev. B}
  \textbf{\bibinfo{volume}{44}}, \bibinfo{pages}{13799} (\bibinfo{year}{1991}).

\bibitem[{\citenamefont{Brocks et~al.}(1993)\citenamefont{Brocks, Kelly, and
  Car}}]{Brocks93AA}
\bibinfo{author}{\bibfnamefont{G.}~\bibnamefont{Brocks}},
  \bibinfo{author}{\bibfnamefont{P.~J.} \bibnamefont{Kelly}}, \bibnamefont{and}
  \bibinfo{author}{\bibfnamefont{R.}~\bibnamefont{Car}},
  \bibinfo{journal}{Phys. Rev. Lett.} \textbf{\bibinfo{volume}{70}},
  \bibinfo{pages}{2786} (\bibinfo{year}{1993}).

\bibitem[{\citenamefont{Bowler}(2004)}]{Bowler04AA}
\bibinfo{author}{\bibfnamefont{D.~R.} \bibnamefont{Bowler}},
  \bibinfo{journal}{J. Phys.: Condens. Matter} \textbf{\bibinfo{volume}{16}},
  \bibinfo{pages}{R721} (\bibinfo{year}{2004}).

\bibitem[{\citenamefont{Liu and Reinke}(2008)}]{Liu08AA}
\bibinfo{author}{\bibfnamefont{H.}~\bibnamefont{Liu}} \bibnamefont{and}
  \bibinfo{author}{\bibfnamefont{P.}~\bibnamefont{Reinke}},
  \bibinfo{journal}{Surf. Sci.} \textbf{\bibinfo{volume}{602}},
  \bibinfo{pages}{986} (\bibinfo{year}{2008}).

\bibitem[{\citenamefont{Sena and Bowler}(2011)}]{Sena11AA}
\bibinfo{author}{\bibfnamefont{A.~M.~P.} \bibnamefont{Sena}} \bibnamefont{and}
  \bibinfo{author}{\bibfnamefont{D.~R.} \bibnamefont{Bowler}},
  \bibinfo{journal}{J. Phys.: Condens. Matter} \textbf{\bibinfo{volume}{23}},
  \bibinfo{pages}{305003} (\bibinfo{year}{2011}).

\bibitem[{\citenamefont{Wang et~al.}(2010)\citenamefont{Wang, Chen, Wang, and
  Kawazoe}}]{Wang10AA}
\bibinfo{author}{\bibfnamefont{J.~T.} \bibnamefont{Wang}},
  \bibinfo{author}{\bibfnamefont{C.}~\bibnamefont{Chen}},
  \bibinfo{author}{\bibfnamefont{E.}~\bibnamefont{Wang}}, \bibnamefont{and}
  \bibinfo{author}{\bibfnamefont{Y.}~\bibnamefont{Kawazoe}},
  \bibinfo{journal}{Phys. Rev. Lett.} \textbf{\bibinfo{volume}{105}},
  \bibinfo{pages}{116102} (\bibinfo{year}{2010}).

\bibitem[{\citenamefont{Niu and Wang}(2011)}]{Niu11AA}
\bibinfo{author}{\bibfnamefont{C.~Y.} \bibnamefont{Niu}} \bibnamefont{and}
  \bibinfo{author}{\bibfnamefont{J.~T.} \bibnamefont{Wang}},
  \bibinfo{journal}{Solid State Commun.} \textbf{\bibinfo{volume}{152}},
  \bibinfo{pages}{127} (\bibinfo{year}{2011}).

\bibitem[{\citenamefont{Cho and Joannopoulos}(1993)}]{Cho93AA}
\bibinfo{author}{\bibfnamefont{K.}~\bibnamefont{Cho}} \bibnamefont{and}
  \bibinfo{author}{\bibfnamefont{J.~D.} \bibnamefont{Joannopoulos}},
  \bibinfo{journal}{Phys. Rev. Lett.} \textbf{\bibinfo{volume}{71}},
  \bibinfo{pages}{1387} (\bibinfo{year}{1993}).

\bibitem[{\citenamefont{Qin and Lagally}(1997)}]{Qin97AA}
\bibinfo{author}{\bibfnamefont{X.~R.} \bibnamefont{Qin}} \bibnamefont{and}
  \bibinfo{author}{\bibfnamefont{M.~G.} \bibnamefont{Lagally}},
  \bibinfo{journal}{Science} \textbf{\bibinfo{volume}{278}},
  \bibinfo{pages}{1444} (\bibinfo{year}{1997}).

\bibitem[{\citenamefont{Dong et~al.}(2001)\citenamefont{Dong, Fujita, and
  Nejoh}}]{Dong01AA}
\bibinfo{author}{\bibfnamefont{Z.~C.} \bibnamefont{Dong}},
  \bibinfo{author}{\bibfnamefont{D.}~\bibnamefont{Fujita}}, \bibnamefont{and}
  \bibinfo{author}{\bibfnamefont{H.}~\bibnamefont{Nejoh}},
  \bibinfo{journal}{Phys. Rev. B} \textbf{\bibinfo{volume}{63}},
  \bibinfo{pages}{115402} (\bibinfo{year}{2001}).

\bibitem[{\citenamefont{Hortamani et~al.}(2006)\citenamefont{Hortamani, Wu,
  Kratzer, and Scheffler}}]{Hortamani06AA}
\bibinfo{author}{\bibfnamefont{M.}~\bibnamefont{Hortamani}},
  \bibinfo{author}{\bibfnamefont{H.}~\bibnamefont{Wu}},
  \bibinfo{author}{\bibfnamefont{P.}~\bibnamefont{Kratzer}}, \bibnamefont{and}
  \bibinfo{author}{\bibfnamefont{M.}~\bibnamefont{Scheffler}},
  \bibinfo{journal}{Phys. Rev. B} \textbf{\bibinfo{volume}{74}},
  \bibinfo{pages}{205305} (\bibinfo{year}{2006}).

\bibitem[{\citenamefont{Hohenberg and Kohn}(1964)}]{Hohenberg64AA}
\bibinfo{author}{\bibfnamefont{P.}~\bibnamefont{Hohenberg}} \bibnamefont{and}
  \bibinfo{author}{\bibfnamefont{W.}~\bibnamefont{Kohn}},
  \bibinfo{journal}{Phys. Rev.} \textbf{\bibinfo{volume}{136}},
  \bibinfo{pages}{B864} (\bibinfo{year}{1964}).

\bibitem[{\citenamefont{Kohn and Sham}(1965)}]{Kohn65AA}
\bibinfo{author}{\bibfnamefont{W.}~\bibnamefont{Kohn}} \bibnamefont{and}
  \bibinfo{author}{\bibfnamefont{L.~J.} \bibnamefont{Sham}},
  \bibinfo{journal}{Phys. Rev.} \textbf{\bibinfo{volume}{140}},
  \bibinfo{pages}{A1133} (\bibinfo{year}{1965}).

\bibitem[{_cp()}]{_cpmd_2010}
\bibinfo{note}{{CPMD,} Copyright {IBM} Corp 1990-2012, Copyright {MPI} f\"ur
  Festk\"orperforschung Stuttgart 1997-2001.},
  \urlprefix\url{http://www.cpmd.org/}.

\bibitem[{\citenamefont{Perdew et~al.}(1996)\citenamefont{Perdew, Burke, and
  Ernzerhof}}]{Perdew96AA}
\bibinfo{author}{\bibfnamefont{J.~P.} \bibnamefont{Perdew}},
  \bibinfo{author}{\bibfnamefont{K.}~\bibnamefont{Burke}}, \bibnamefont{and}
  \bibinfo{author}{\bibfnamefont{M.}~\bibnamefont{Ernzerhof}},
  \bibinfo{journal}{Phys. Rev. Lett.} \textbf{\bibinfo{volume}{77}},
  \bibinfo{pages}{3865} (\bibinfo{year}{1996}).

\bibitem[{\citenamefont{Hamann}(1989)}]{Hamann89AA}
\bibinfo{author}{\bibfnamefont{D.~R.} \bibnamefont{Hamann}},
  \bibinfo{journal}{Phys. Rev. B} \textbf{\bibinfo{volume}{40}},
  \bibinfo{pages}{2980} (\bibinfo{year}{1989}).

\bibitem[{\citenamefont{Kleinman and Bylander}(1982)}]{Kleinman82AA}
\bibinfo{author}{\bibfnamefont{L.}~\bibnamefont{Kleinman}} \bibnamefont{and}
  \bibinfo{author}{\bibfnamefont{D.~M.} \bibnamefont{Bylander}},
  \bibinfo{journal}{Phys. Rev. Lett.} \textbf{\bibinfo{volume}{48}},
  \bibinfo{pages}{1425} (\bibinfo{year}{1982}).

\bibitem[{\citenamefont{Ramstad et~al.}(1995)\citenamefont{Ramstad, Brocks, and
  Kelly}}]{Ramstad95AA}
\bibinfo{author}{\bibfnamefont{A.}~\bibnamefont{Ramstad}},
  \bibinfo{author}{\bibfnamefont{G.}~\bibnamefont{Brocks}}, \bibnamefont{and}
  \bibinfo{author}{\bibfnamefont{P.~J.} \bibnamefont{Kelly}},
  \bibinfo{journal}{Phys. Rev. B} \textbf{\bibinfo{volume}{51}},
  \bibinfo{pages}{14504} (\bibinfo{year}{1995}).

\end{thebibliography}
\end{document}